\documentclass{emulateapj}

\usepackage{verbatim}
\newcommand{\unit}[1]{\,{\rm #1}}

\begin{document}

\title{Density Probability Distribution Functions in Supersonic Hydrodynamic
and MHD Turbulence}
\author{M. Nicole Lemaster\altaffilmark{1} and James M. Stone}
\affil{Department of Astrophysical Sciences, Princeton University,
Princeton, NJ 08544}
\altaffiltext{1}{\email{Lemaster@astro.princeton.edu}}

\shorttitle{Density PDFs in Supersonic Turbulence}
\shortauthors{Lemaster \& Stone}

\begin{abstract}

We study the probability distribution function (PDF)
of the mass density in simulations of supersonic turbulence with
properties appropriate for molecular clouds.  For this study we use
Athena, a new higher-order Godunov code.
We find there are surprisingly similar relationships between the mean
of the time-averaged PDF and the turbulent Mach number for driven
hydrodynamic and strong-field MHD turbulence.  There is, however, a
large scatter about these relations, indicating a high level of
temporal and spatial variability in the PDF.  Thus, the PDF of the
mass density is unlikely to be a good measure of magnetic field strength.
We also find that the PDF of decaying MHD turbulence deviates from the
mean-Mach relation found in the driven case.  This implies that the
instantaneous Mach number alone is not enough to determine the
statistical properties of turbulence that is out of equilibrium.
The scatter about the mean-Mach relation for driven turbulence, along
with the large departure of decaying turbulence PDFs from those of
driven turbulence, may illuminate one factor contributing to the large
observed cloud-to-cloud variation in the star formation rate per solar
mass.

\end{abstract}


\section{Introduction}


The mechanism that determines the star formation rate (SFR) within a
molecular cloud (MC) is not well understood.  Observations show that
the SFR per solar mass, as measured by the ratio of CO to IR
luminosity, varies by as much as 2-3 orders of magnitude from region
to region; instead Gao \& Solomon (2004) found that HCN
emission, which measures molecular gas at much higher density
($\gtrsim 3 \times 10^4 \unit{cm^{-3}}$), is the better indicator of
star formation.  This suggests that the amount of mass to reach high
density is the key factor in determining the SFR per solar mass within
a cloud (McKee \& Ostriker 2007).  Observed non-thermal line widths in
MCs (Falgarone \& Philips 1990) indicate that supersonic turbulence
may be responsible for creating the high density contrasts that lead
to clump formation.


Although magnetic fields have been shown not to lengthen appreciably
the decay timescale of supersonic turbulence (Stone, Ostriker, \&
Gammie 1998, hereafter S98; Mac Low 1999), they do create anisotropy
in the structures within the medium (Vestuto et al. 2003).
Observations using Zeeman splitting, such as those described
by Crutcher (1999), have found magnetic fields in some clouds strong
enough (e.g. $\beta \approx 0.04$) that they cannot be neglected.
The magnetic field within a MC is typically difficult to detect,
motivating the determination of new diagnostics (e.g. Heyer et al.
2008).


The probability distribution function (PDF) of the logarithm of mass
density can be used to quantify the amount of material within a
turbulent medium that has a given density.  As the density is likely
to have a strong impact on star formation, many groups have
investigated the properties of such PDFs.  Padoan et al.
(1997, hereafter P97), for example, conducted simulations of driven
hydrodynamic turbulence, determining a relation between the mean of
the PDF and the Mach number.  Most of the mass in the simulated clouds
was found to be in only a small fraction of the volume, with the width
of the approximately Gaussian PDF increasing with Mach
number in a predictable way.  Ostriker, Stone, \& Gammie (2001)
investigated the effect of magnetic field strength on the mean-Mach
relation for decaying turbulence, finding that the fast magnetosonic
Mach number can be used to predict only a lower limit on the width of
the distribution.


Recently, we have undertaken a comprehensive study of the properties
of supersonic turbulence (Lemaster \& Stone 2008, hereafter Paper I)
with Athena, a new higher-order Godunov
method (Stone et al. 2008).  This study represents one of the first
applications of Godunov methods to the study of supersonic MHD
turbulence, and therefore represents an important test of previous results.
In this letter, we
investigate the effects of magnetic field strength and Mach number on
the PDF.  We survey a much larger range of Mach number and field
strength, and at a higher numerical resolution, than has been used in
previous studies.  This provides much better statistics and allows us
to constrain the form of the relations more tightly.  We describe the
numerical methods used to conduct our simulations in
\S\ref{sec:methods}, explain the focus of our analysis in
\S\ref{sec:pdfs}, and present the results in \S\ref{sec:results}.
Finally, we summarize in \S\ref{sec:concl}.

\section{Numerical Methods}\label{sec:methods}


The simulations we present here were conducted at a resolution of
$512^3$ with Athena (Gardiner \& Stone 2005, 2008; Stone et al.
2008; Stone \& Gardiner 2008), a new higher-order Godunov code that
exactly conserves mass, momentum, and magnetic flux.  We solve the
equations of ideal isothermal MHD,
\begin{equation}
\frac{\partial \rho}{\partial t} + {\bf \nabla} \cdot (\rho {\bf v}) = 0,
\end{equation}
\begin{equation}
\frac{\partial \rho {\bf v}}{\partial t} +
{\bf \nabla} \cdot (\rho {\bf vv} - { \bf BB} + P + B^2/2) = 0,
\end{equation}
and
\begin{equation}
\frac{\partial {\bf B}}{\partial t} =
{\bf \nabla} \times ({\bf v} \times {\bf B}),
\end{equation}
where $c_s = 1$ and $P = c_s^2 \rho$ are the isothermal sound speed
and pressure, on a three-dimensional periodic Cartesian grid of
length $L = 1$.  We use an
approximate nonlinear Riemann solver (HLLD; Miyoshi \& Kusano 2005) for
our MHD runs and an exact nonlinear Riemann solver for our hydrodynamic
runs.  We integrate our simulations for roughly 4 dynamical times
beyond the saturation time using a directionally-unsplit van Leer
scheme (Stone \& Gardiner 2008).  Here we have defined the dynamical
time to be $t_{\rm dyn} = L/(2\mathcal{M})$, where
$\mathcal{M} \equiv \sigma_v/c_s$ is the Mach number and
$\sigma_v = \langle v^2 \rho/\bar{\rho} \rangle^{1/2}$ is the velocity
dispersion of the gas, calculated using a mass-weighted average.
For further details on our numerical methods as applied to
turbulence, see Paper I.


We drive our turbulence here in a manner very similar to S98.
We initialize a uniform, stationary ambient medium
with density $\bar{\rho} = 1$ and magnetic field parallel to the
$x$-axis whose amplitude $B_0$ is fixed by the value of
$\beta = 2c_s^2\bar{\rho}/B_0^2$.  We apply divergence-free
velocity perturbations before every time step, following a Gaussian
random distribution peaked at $k_{\rm pk}L/2\pi = 2$.  Before applying
the perturbations, we shift them such that no net momentum will be
added to the grid.  We also normalize them to give the desired energy
injection rate, $\dot{E}/\bar{\rho}L^2c_s^3$.  For our decaying runs,
we begin with a snapshot of fully-developed turbulence from a driven
run and allow it to evolve without further energy injection.


We investigate strong-field MHD ($\beta = 0.02$) as well as
hydrodynamic ($\beta = \infty$) turbulence.  The magnetic fields we
use in our simulations correspond to physical values of
$B = 2.0 \unit{\mu G} \beta^{-1/2} (T/10 \unit{K})^{1/2}
(n_{H_2}/10^2 \unit{cm^{-3}})^{1/2}$, where $T$ is the temperature and
$n_{H_2}$ is the number density of molecular hydrogen.  Our
simulations are scale-free, allowing them to be scaled to any set of
physical parameters using appropriate choices of $\bar{\rho}$, $c_s$,
and $L$.  Utilizing the same values given in S98, i.e.
$L = 2 \unit{pc}$, $n_{H_2} = 10^3 \unit{cm^{-3}}$, and
$T = 10 \unit{K}$, yields an energy injection rate of
$\dot{E} = 0.2 L_\odot$ with magnetic field strength
$B = 44 \unit{\mu G}$.

\section{Probability Distribution Functions}\label{sec:pdfs}


Turbulence in molecular clouds causes converging flows where the gas
can be compressed to very high densities.  The PDF of the density
tells us the fraction of the mass or volume within a cloud that
obtains a given density.  Since self-gravitating clumps can form in
the high-density regions, understanding PDFs is critical for gaining
insight into the stellar IFM and SFR.
If compression and rarefaction events in the turbulent gas within a
molecular cloud are spatially and temporally independent, the PDF of
density will have a log-normal distribution (Passot \&
Vazquez-Semadeni 1998).  The PDF of the logarithm of density, then,
will have a normal distribution given by
\begin{equation}
f(y) dy = \frac{1}{\sqrt{2\pi\sigma^2}}
\exp \Big[ \frac{-(y-\mu)^2}{2\sigma^2} \Big] dy,
\end{equation}
where $y \equiv \ln (\rho/\bar{\rho})$, $\mu$ is the mean of the
distribution, and $\sigma^2$ is the dispersion, with
$|\mu| = \sigma^2/2$.

Our goal in this Letter is to analyze the relationship between the
mean of the distribution, which represents the median density within
the cloud, and the Mach number.  We have investigated this
mean-Mach relation over a range of Mach numbers
$1.2 \le \mathcal{M} \le 7.0$ for both driven hydrodynamic and
strong-field MHD turbulence.  We found that a resolution of $512^3$
gave smoother-looking PDFs that could be fitted more accurately than
those from $256^3$ simulations, justifying the computational expense.
Higher resolution also allows us to study scatter in the PDF in
sub-volumes of the domain.  We show in Paper I that the simulations
have converged by this resolution; thus, increasing it further is
unlikely to affect the results.  To minimize the influence of
intermittency, we time-average the PDFs obtained from seven snapshots
in the saturated state spanning almost 3 dynamical times before
fitting them.

Since the tails of the PDFs will deviate from normal form due to the
effects of intermittency, we fit only bins with values of at least
$10\%$ of the peak value.  We perform a Levenberg-Marquardt
least-squares fit with uniform weighting.  Once we have obtained the
mean, $\mu$, of the best-fit distribution, we plot it against a
function of Mach number,
$\xi(\mathcal{M}) = \ln[1+\alpha\mathcal{M}^2]$.  With the appropriate
choice of $\alpha$, we can obtain a linear relation between the PDF
mean and this function $\xi(\mathcal{M})$.  We note that Kowal et al.
(2007) have recently shown that higher order statistics can also
provide insight into the properties of supersonic turbulence, however
in this Letter we will focus only on the density PDF.

Observations show a wide range of star formation rates for different
clouds.  To infer from our simulations the level of cloud-to-cloud
variation we would likely observe, we also investigate the mean-Mach
relation for regions of size comparable to the driving scale.  To do
this, we divide our computational domain into eight equal sub-domains,
each of resolution $256^3$.  We compute the PDF in each of these
sub-boxes (which we will refer to as sub-PDFs) individually and plot
their means against the Mach number within that sub-box.
We do not time-average our results, yielding 56 mean-Mach pairs for
each run.  Although the snapshots are at intervals of just under half
a dynamical time on the global scale, the interval between snapshots
is closer to a flow crossing time on the scale of the sub-boxes,
making the snapshots sufficiently uncorrelated for our analysis.
Substantial scatter in the values within a run might help explain the
observed cloud-to-cloud variation in the star formation rate in
molecular clouds.

\section{Results}\label{sec:results}


The time-averaged PDFs over the full domain are very smooth and
approximate Gaussians, particularly in the hydro case.
Although we have not plotted such a PDF in this Letter, a similar
example can be seen in Figure 3 of Kritsuk et al. (2007).  Presented
in Table \ref{tab:runs} are our PDF statistics.  The first two
columns, respectively, specify the magnetic $\beta$ and energy
injection rate.  The final six columns are determined from the
sub-volumes and describe the error bars in the figures to follow.
The values $\sigma_{V,S}$ and $\sigma_{M,S}$ in the table are the
standard deviation of the means determined from the PDFs, not the
width of the PDFs themselves.

\begin{deluxetable}{cccccccc}
\tablecolumns{8}
\tablewidth{0pc}
\tablecaption{Driven Turbulence at $512^3$\label{tab:runs}}
\tablehead{\colhead{$\beta$} & \colhead{$\dot{E}$}
& \colhead{$\mathcal{M}$} & \colhead{$\sigma_{\mathcal{M}}$}
& \colhead{$\mu_V$} & \colhead{$\sigma_{V,S}$\tablenotemark{a}}
& \colhead{$\mu_M$} & \colhead{$\sigma_{M,S}$\tablenotemark{a}}}
\startdata
$\infty$ & 500    & 6.8 & 0.5  & -1.1  & 0.13 & 0.94 & 0.09 \\
$\infty$ & 187.5  & 5.0 & 0.4  & -0.87 & 0.12 & 0.76 & 0.10 \\
$\infty$ &  70    & 3.6 & 0.3  & -0.65 & 0.08 & 0.57 & 0.07 \\
$\infty$ &  20    & 2.4 & 0.1  & -0.40 & 0.05 & 0.35 & 0.05 \\
$\infty$ &   1.75 & 1.2 & 0.05 & -0.10 & 0.02 & 0.09 & 0.01 \\
\hline
  0.02   & 500    & 6.7 & 0.5  & -0.89 & 0.12 & 0.84 & 0.11 \\
  0.02   & 187.5  & 4.9 & 0.3  & -0.71 & 0.12 & 0.66 & 0.11 \\
  0.02   &  70    & 3.6 & 0.2  & -0.56 & 0.09 & 0.50 & 0.09 \\
  0.02   &  20    & 2.5 & 0.1  & -0.34 & 0.05 & 0.30 & 0.05 \\
  0.02   &   1.75 & 1.2 & 0.06 & -0.12 & 0.03 & 0.11 & 0.03
\enddata
\tablenotetext{a}{$\sigma_{V,S}$ and $\sigma_{M,S}$ are the standard deviation of $\mu_V$ and $\mu_M$, respectively.}
\end{deluxetable}

\subsection{Driven Hydrodynamic Turbulence}\label{sec:hydpdf}

\begin{figure}
\epsscale{1.0}
\plotone{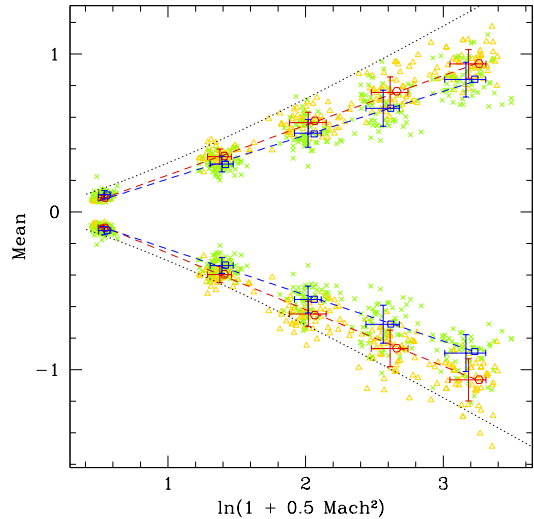}
\figcaption{
PDF mean versus $\xi(\mathcal{M})$ for driven hydro (red hexagons)
and strong-field MHD (blue squares) turbulence.  Only time-averaged
PDFs over the full domain are used to fit the mean-Mach relations
(hydro: red dashed; MHD: blue dashed).  Also shown are the
instantaneous values taken from each of the eight driving-scale
sub-boxes (hydro: gold triangles; MHD: green crosses) and the
1$\sigma$ error bars on those points (hydro: red; MHD: blue).
For comparison, the relation from P97 is also shown (black dotted).
\label{fig:pdfreln}}
\end{figure}

Figure \ref{fig:pdfreln} shows the mean-Mach relation found from
time-averaged PDFs over the full domain for driven hydro and driven MHD
turbulence.  We find that a value of $\alpha = 0.5$ in the function
$\xi(\mathcal{M})$ gives the best linear relations for the hydro case.
For the volume fraction we find
\begin{equation}
\mu_V = -0.36 \ln[1+0.5\mathcal{M}^2] + 0.10,
\end{equation}
while for the mass fraction we find
\begin{equation}
\mu_M = 0.32 \ln[1+0.5\mathcal{M}^2] - 0.10.
\end{equation}
The rms residuals for these fits are $8.9 \times 10^{-3}$ and
$6.5 \times 10^{-3}$, respectively.
Because the density fluctuations in subsonic turbulence are not
produced by shocks, we have no reason to expect these relations to
approach zero with Mach number.
The mean-Mach pairs from the time-averaged PDFs fall very close to
these relations.  Over the full range of Mach numbers tested, however,
the time-averaged means are smaller than those found by P97
($\mu_{V,M} = \mp0.5 \ln[1+0.25\mathcal{M}^2]$).
To determine the magnitude of the effect that the driving may have had
on the relations, we also compare values determined from hydrodynamic
turbulence with $k_{\rm pk}L/2\pi = 4$ (not shown), finding that these
points fall very close to the $k_{\rm pk}L/2\pi = 2$ relations as well.

Also shown in the figure are the instantaneous mean-Mach pairs found
from each of the eight sub-boxes.  The scatter in these points,
illustrated by 1$\sigma$ error bars on the plot, is quite
significant.  The ensemble average of the 56 points for each run falls
at a slightly lower Mach number than the value found from the
time-average over the full domain since the Mach number typically
decreases on smaller scales.  These averages differ slightly from
those computed from time-averaged full-domain PDFs, although they
still fall close to the relations found.  The 1$\sigma$ scatter in the
Mach number is $4\%$--$8\%$, while in the sub-PDF means it is
$16\%$--$17\%$ for the lowest Mach number run and $10\%$--$14\%$ for the
remaining runs.  This scatter puts some of the instantaneous sub-PDF
values in the vicinity of the P97 relation.

The scatter in $\mu_V$ and $\mu_M$ can be converted to scatter in the
median density within the medium.  For the ratio of the median to mean
density within a cloud, $\tilde{\rho}$, we can define $\tilde{\rho}_+$
and $\tilde{\rho}_-$ for clouds where $\mu$ is 1$\sigma$ above or below
the mean.  The ratios
$\tilde{\rho}_{V,+}/\tilde{\rho}_{V,-}=\exp(2\sigma_{V,S})$ and
$\tilde{\rho}_{M,+}/\tilde{\rho}_{M,-}=\exp(2\sigma_{M,S})$ then
generally increase with Mach number, ranging from 1.0 to 1.3.

\subsection{Driven MHD Turbulence}\label{sec:mhdpdf}


Figure \ref{fig:pdfreln} also includes the mean-Mach relation for
driven strong-field MHD turbulence, where we continue to use
$\alpha = 0.5$.  The mean-Mach pairs from the time-averaged PDFs
again fall very close to these relations,
\begin{equation}
\mu_V = -0.29 \ln[1+0.5\mathcal{M}^2] - 0.06
\end{equation}
for the volume fraction and
\begin{equation}
\mu_M = 0.28 \ln[1+0.5\mathcal{M}^2] + 0.07
\end{equation}
for the mass fraction,
still yielding means smaller than those found by P97.
The rms residuals for these fits are $1.1 \times 10^{-2}$ and
$1.6 \times 10^{-2}$, respectively.

The instantaneous mean-Mach pairs found from sub-boxes have more
scatter with a strong magnetic field than they did in the purely
hydrodynamic case.  The 1$\sigma$ scatter in the Mach number is
$5\%$--$8\%$, while for the sub-PDF means it is $22\%$--$24\%$ for the
lowest Mach number run and $13\%$--$18\%$ for the remaining runs.
Converting this to scatter in the median density, we find that the
ratios $\tilde{\rho}_{V,+}/\tilde{\rho}_{V,-}$ and
$\tilde{\rho}_{M,+}/\tilde{\rho}_{M,-}$ range from 1.1 to 1.3.  This
again puts some of the instantaneous sub-PDF values in the vicinity of
the P97 mean-Mach relation.

The time-averaged mean-Mach relations found for hydro and MHD differ,
as one should expect due to differences in the shock-jump conditions.
However, the sub-PDF values overlap substantially, making them
difficult to distinguish observationally.  The time-averaged relations
fall a bit less than 1$\sigma$ (computed from the sub-PDF values)
apart.

\subsection{Decaying MHD Turbulence}\label{sec:decpdf}

\begin{figure}
\epsscale{1.0}
\plotone{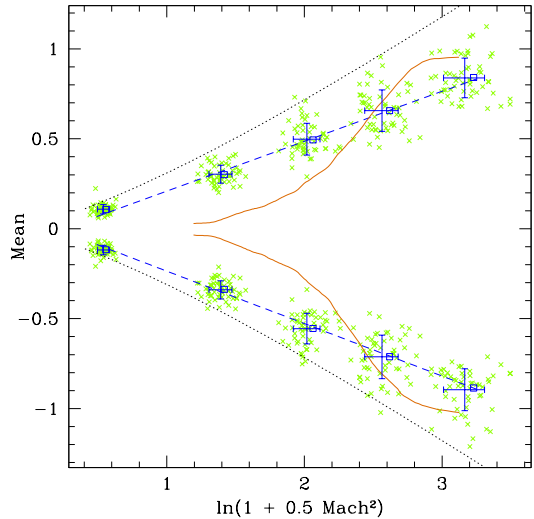}
\figcaption{
PDF mean versus $\xi(\mathcal{M})$ for decaying MHD turbulence
(orange solid) as it evolves over time (right to left), contrasted with
the time-averaged driven values (blue squares) and relation found in
\S\ref{sec:mhdpdf} (blue dashed).  The decaying run does not follow
the relation found for driven MHD, even considering the large scatter
(error bars and green crosses) found in the driven sub-PDFs.  For
comparison, the hydro relation from P97 is also shown (black dotted).
\label{fig:decreln}}
\end{figure}

As it seems likely that molecular clouds are transient entities,
it may be more appropriate to study the PDF of decaying turbulence.
In Figure \ref{fig:decreln}, we show the evolution of the PDF in a
decaying strong-field MHD turbulence run, initialized from a snapshot
of fully-developed turbulence from our highest Mach number driven run.
Although this snapshot has a full-domain PDF mean roughly 1$\sigma$
more extreme than the time-averaged driven relations, this should not
affect the results.  At first the small change in mean as the Mach number
decreases causes a shallower slope than that of the driven relation.
Once the mean begins to change appreciably, however, the slope becomes
much steeper, crossing the driven relation at roughly $\mathcal{M}
= 4.5$.  Although the slope shallows as low Mach numbers are reached,
it does so only as the means become very small.

The evolution of this decaying run does not parallel or
asymptotically approach the driven relation at the same magnetic
$\beta$.  Contrary to what was found by Passot \& Vazquez-Semadeni
(1998) for one-dimensional polytropic gas, then, we find that the
instantaneous Mach number is insufficient to describe the
properties of the turbulent medium when the gas is not in a
statistically steady state.  This is consistent with the findings of
Ostriker et al. (2001) for the PDFs of decaying turbulence.  If MCs
contain decaying turbulence, which seems likely (Heitsch et
al. 2006), the PDF may be ``out of equilibrium'', making relations
obtained from steady state (driven) turbulence inapplicable.

\section{Summary}\label{sec:concl}


For both supersonic hydrodynamic and strong-field MHD turbulence, we
have found a one-to-one correspondence between the mean of the
time-averaged PDF and the Mach number.  The mean-Mach pairs
used to fit these relations have very small residuals; however, the
mean of the PDF at any given Mach number, in both the hydro and MHD
cases, is smaller than was found by P97 for the purely hydrodynamic
case.  Although there is substantial scatter of the mean-Mach pairs
computed from instantaneous sub-volumes, the ensemble average of these
values still falls close to the time-averaged global relation.  The
scatter puts a small fraction of the instantaneous values in the
vicinity of the P97 relation.

Although the relations found for hydro and MHD differ, the scatter
about the mean relation of the instantaneous sub-PDF values creates
substantial overlap between the two.  Since the MHD relation gives means
that are typically smaller than the corresponding hydrodynamic values
by only 1$\sigma$, it will be very difficult to distinguish between
the two observationally.  We have also found that PDFs in decaying MHD
turbulence differ from those of driven MHD turbulence at the same
magnetic $\beta$.  It would seem that the instantaneous Mach number
alone does not adequately describe the statistical state of the
turbulent gas when not in equilibrium.  Again, however, there is
substantial overlap between the equilibrium and non-equilibrium values,
preventing this diagnostic from being used to distinguish between the
two.

The scatter we have found about the mean-Mach relation may help
explain the large variation in the observed SFR per solar mass in
molecular clouds.  Since MCs are likely to be transient entities,
relations found from driven turbulence may not even be applicable to
real clouds.  If large departures from the mean-Mach relation are in
fact linked to the large variation in the SFR per solar mass, this may
indicate that turbulence in MCs is indeed decaying rather than forced.

\acknowledgements

We thank Eve Ostriker for very productive discussion.  Simulations
were performed on the IBM Blue Gene at Princeton and on computational
facilities supported by NSF grants AST-0216105 and AST-0722479.

\end{document}